\documentclass[runningheads,envcountsect]{llncs}

\usepackage{amsmath,amssymb}
\usepackage{enumitem}
\setlist[enumerate]{leftmargin=*}
\usepackage[expansion=false]{microtype}
\usepackage{hyperref}

\newenvironment{proofof}[1]{\par\addvspace{6pt}\noindent\textit{Proof of #1.}\enspace\ignorespaces}{\par\addvspace{6pt}}
\spnewtheorem{assumption}{Assumption}{\bfseries}{\itshape}
\makeatletter\@addtoreset{assumption}{section}\makeatother

\begin{document}

\title{Exchange Rate Determination for Cryptocurrency Mergers:
A Formal Framework}
\titlerunning{Exchange Rate Determination for Cryptocurrency Mergers}

\author{Massimiliano Sala\inst{1} \and Daniela Visetti\inst{2}}
\authorrunning{M. Sala and D. Visetti}

\institute{Department of Mathematics, University of Trento, Italy
\and
Department of Economics, Management and Statistics, University of Milano-Bicocca, Italy}

\maketitle

\begin{abstract}
Many of the thousands of existing cryptocurrencies suffer from declining adoption, low liquidity and weak security, and merging two of them into a single ecosystem is a natural alternative to abandonment. No rigorous framework exists, however, for determining a fair exchange rate in such a merger. Unlike that of a corporation, the value of a cryptocurrency is driven by network effects, so the exchange rate itself influences the value of the merged asset. We extend the exchange-ratio framework of Mainini, Moretto and Visetti \cite{Mainini2024} by making the merged value endogenous: the exchange rate determines community migration, migration determines (up to user overlap) the post-merger network state, and an axiomatised valuation function assigns a value to that state. The resulting synergy has no predetermined sign, and the bounds of the bargaining region become self-referential in the exchange rate. We prove that no exchange rate at which the merger destroys value preserves wealth, and that the pre-merger price ratio preserves wealth exactly when the merger does not destroy value at that rate. In that case, under explicit threshold conditions, the price ratio satisfies all five of our fairness conditions --- wealth, adoption, security, governance and liquidity preservation --- and, under strict versions of these conditions, the admissible set contains a nondegenerate interval around it, the wealth-admissible set one of explicit width. The network-value laws proposed in the literature, from the linear law to Metcalfe's and its generalised power-law forms, share a convexity property that yields the Lipschitz control behind these estimates. Liquidity emerges as the only non-wealth condition that can disconnect the admissible set.

\end{abstract}

\section{Introduction}

Thousands of cryptocurrencies have been launched during the last decade. While a small number have achieved significant adoption, the overwhelming majority remain characterized by low trading volumes, fragmented communities and decreasing development activity.

Many projects ultimately become economically unsustainable despite possessing useful technology. A natural alternative to project abandonment is the voluntary merger of two or more blockchain ecosystems into a single cryptocurrency. Examples already exist in practice: token migrations, blockchain consolidations and protocol upgrades frequently require holders of one cryptocurrency to exchange their tokens for those of another project. These conversions are generally negotiated on an ad hoc basis, with little theoretical justification for the adopted conversion ratio.

This raises a natural question:

\begin{quote}
\emph{How should two cryptocurrencies determine a fair exchange rate before merging?}
\end{quote}

Traditional corporate finance addresses an analogous problem through exchange-ratio determination in stock-for-stock mergers. Mainini, Moretto and Visetti \cite{Mainini2024} extend, to a stochastic setting, the deterministic model of Larson and Gonedes \cite{Larson1969} and Yagil \cite{Yagil1987} (together referred to, following \cite{Mainini2024}, as the LG-Y setting), producing an explicit \emph{bargaining region} of acceptable exchange ratios. Section~\ref{sec:mainini} recalls this framework in full, using exactly the notation of \cite{Mainini2024}, since our extension reuses it rather than inventing parallel symbols.

Cryptocurrencies differ fundamentally from corporations in one respect that the LG-Y/Mainini setting does not need to address: a cryptocurrency derives much of its value from network participation, so the exchange ratio chosen for the merger directly influences the future value of the merged cryptocurrency by affecting user migration and ecosystem growth. In the notation recalled below, the merged entity's equity value $\mu_M$ is treated by \cite{Mainini2024} as an exogenous quantity (parametrized by an exogenous synergy $s\geq 0$); our contribution is to make $\mu_M$ endogenous, i.e.\ a function of $r$ itself, through a network-driven synergy term $s(r)$ that we construct explicitly in Section~\ref{sec:setup}. This is the central mathematical obstruction addressed in this paper.

The remainder of the paper is organised as follows. Section~\ref{sec:mainini} recalls the Mainini et al.\ framework and its notation. Section~\ref{sec:related} reviews the 
cryptocurrency economics literature this paper draws on and explains how the merger problem relates to, and differs from, each strand. Section~\ref{sec:different} explains precisely 
where the cryptocurrency setting departs from the corporate framework. Section~\ref{sec:setup} introduces the state-space description of a cryptocurrency, an axiomatised network 
valuation function, and the resulting endogenous synergy $s(r)$. Section~\ref{sec:fairness} defines the fairness conditions as explicit subsets of the exchange-rate line, one of 
which (wealth preservation) is exactly the bargaining region of Section~\ref{sec:mainini}. Section~\ref{sec:static} determines the structure of the four non-wealth sets. 
Section~\ref{subsec:selfref} 
treats the self-referential wealth condition and establishes existence of self-consistent boundary rates. The value-creation dichotomy is proved. Section~\ref{sec:r*} shows 
that the neutral rate $r_*$ is a common point of all five conditions under a checkable threshold condition, so that the admissible set is nonempty. 
Section~\ref{sec:mainthm} surveys concrete proposals for the valuation 
function from the network-economics and cryptocurrency literature, isolates their shared structural property (convexity with value zero at zero), and uses it to prove 
explicit width bounds. Section~\ref{sec:conclusions} summarises the results and lists open problems.

\section{A stock-for-stock merger}
\label{sec:mainini}

We recall here, using the notation of \cite{Mainini2024} verbatim, the deterministic part of their framework, since it is what our extension specializes.

Consider a stock-for-stock merger between an acquiring company $A$ and an acquired (target) company $B$, resulting in company $M$. Let $p_i$, $N_i$ and $\mu_i := p_i N_i$ denote, respectively, the stock price, the number of outstanding stocks, and the equity value of company $i \in \{A,B,M\}$. Each stock of $A$ becomes one stock of $M$; each stock of $B$ becomes $r$ stocks of $M$, where $r$ is the exchange ratio (ER). Consequently
\[
N_M = N_A + rN_B.
\]

A relevant reference quantity is
\begin{equation}
\label{eq:price_ratio}
r_* := \frac{p_B}{p_A},
\end{equation}
the price ratio of the two companies' stocks prior to the merger.

Shareholders of $A$ benefit from the merger when $p_M = \mu_M/(N_A+rN_B) \geq \mu_A/N_A = p_A$, equivalently when
\begin{equation}
\label{eq:over_r_mu}
r \leq r_* \cdot \frac{\mu_M - \mu_A}{\mu_B} =: \overline{r}_\mu(\mu_M).
\end{equation}
Shareholders of $B$ benefit when $r\cdot p_M \geq \mu_B/N_B = p_B$, equivalently when
\begin{equation}
\label{eq:under_r_mu}
r \geq r_* \cdot \frac{\mu_A}{\mu_M - \mu_B} =: \underline{r}_\mu(\mu_M).
\end{equation}
Whenever $\mu_M \geq \mu_A + \mu_B$ (i.e.\ the merger creates non-negative synergy), $\underline{r}_\mu(\mu_M) \leq \overline{r}_\mu(\mu_M)$, and the two bounds define the \emph{bargaining region}
\[
\mathcal{BR}_\mu(\mu_M) := [\underline{r}_\mu(\mu_M), \overline{r}_\mu(\mu_M)],
\]
which collapses to the singleton $\{r_*\}$ in the absence of synergy ($\mu_M = \mu_A+\mu_B$). Writing $s \geq 0$ for the expected synergy, so that
\begin{equation}
\label{eq:synergy}
\mu_M = \mu_A + \mu_B + s,
\end{equation}
$\mathcal{BR}_\mu$ is nonempty and nondegenerate exactly when $s > 0$, and \cite{Mainini2024} treats $s$ (equivalently $\mu_M$) as an exogenous parameter, varying it to trace out the whole family of bargaining regions.

\section{Related literature on cryptocurrency economics}
\label{sec:related}

To our knowledge, no prior work addresses exchange-ratio determination for cryptocurrency mergers, but several strands of cryptocurrency economics bear on the ingredients of our model.

\paragraph{Token valuation with endogenous adoption.}
Cong, Li and Wang \cite{CongLiWang2021} price a token by aggregating users' transactional demand, with network externalities creating a feedback between adoption and price. This is the single-platform analogue of our loop $r \to \tau_A,\tau_B \to s(r) \to \mathcal{BR}_\mu(\mu_M(r))$; their model has no exchange ratio between two communities, and our axioms (F1)--(F3) can be read as a static reduced form of their map from user base to value.

\paragraph{Fragility under network effects.}
Sockin and Xiong \cite{SockinXiong2023} show that network effects in participation can make the token market break down, with participation collapsing to zero. The map from announced terms to realised participation then need not be continuous or single-valued, whereas Assumption~\ref{ass:migration} imposes continuous monotone migration, which we regard as a tractable first approximation.

\paragraph{Forks.}
Biais, Bisi\`ere, Bouvard and Casamatta \cite{BiaisEtAl2019} model proof-of-work as a coordination game with equilibria featuring persistent forks. A merger is the inverse operation of a fork: whether an announced $r$ leads to consolidation or to a de facto continued fork is a coordination problem, which our migration functions summarise in reduced form. Contested forks are natural data for calibrating them.

\paragraph{Competition.}
Gandal and Halaburda \cite{GandalHalaburda2016} find winner-take-all dynamics among cryptocurrencies in the early market and weaker concentration later. Mergers are motivated when network effects are strong; when they are weak, $s(r)$ is nearly flat and our framework reduces to the exogenous-synergy setting of \cite{Mainini2024}.

\paragraph{Network-value laws.}
Peterson \cite{Peterson2018} models the value of Bitcoin by Metcalfe's law, and Alabi \cite{Alabi2017} finds that several blockchain networks follow Metcalfe-type laws. Together with the laws of \cite{BriscoeOdlyzkoTilly2006,Wheatley2019}, these supply the user-network components of the valuation functions of Section~\ref{sec:mainthm}.

\section{Why cryptocurrency mergers are different}
\label{sec:different}

In \cite{Mainini2024}, $\mu_A, \mu_B$ are observed market equity values, and $s$ (hence $\mu_M$) is a free exogenous parameter representing the merger's financial synergy — chosen by the analyst to trace the bargaining region, not determined by $r$ itself.

For a cryptocurrency merger this assumption is questionable. The value of a cryptocurrency depends upon several endogenous quantities, including active users, trading liquidity, validator participation, developer activity, and market confidence, all of which may change \emph{after} the exchange rate $r$ is announced: an exchange ratio perceived as unfair may cause one community to reject the merger and decline to migrate, directly reducing $s$. We therefore hypothesize the dependency
\[
r \;\longrightarrow\; \text{user migration} \;\longrightarrow\; \text{network synergy } s \;\longrightarrow\; \text{fair exchange rate},
\]
which closes into a loop: the synergy term that \cite{Mainini2024} treats as exogenous must, for cryptocurrencies, be written as a function $s(r)$, making $\mu_M(r) = \mu_A + \mu_B + s(r)$ endogenous and the bargaining-region bounds $\overline{r}_\mu(\mu_M(r))$, $\underline{r}_\mu(\mu_M(r))$ self-referential in $r$. This observation motivates the remainder of the paper.

\section{Formal setup}
\label{sec:setup}

The purpose of this section is to make the merged network value endogenous to the exchange ratio. We first describe each cryptocurrency by a network state and introduce a valuation function $F$. We then let the exchange ratio determine migration and hence the post-merger state, producing the endogenous value 
$\mu_M(r)=F(\sigma_M)$ and synergy 
$s(r)=\mu_M(r)-\mu_A-\mu_B$. The proofs of all results are given in Appendix~\ref{app:proofs}.

\subsection{State space}

Estimating the value of a cryptocurrency is a hard task, because it depends on many factors.\\
As a first approximation, in this paper we assume that the blockchain value depends only on four parameters, that
live in the following state set $\mathcal D$, with $\mathbb{R}^+$ denoting the 
nonnegative real numbers,
\[
\mathcal D
:=
\mathbb R^+
\times
\mathbb R^+
\times
[0,1]
\times
[0,1],
\]
where, for a state $\sigma=(U,L,S,G)\in \mathcal D$, $U$ is the number of active users, $L$ is a measure of network liquidity, $S \in[0,1]$ is a normalised security or validator-participation score, and $G \in[0,1]$ is a normalised governance-participation score.

The same definitions and measurement conventions are used for both
networks. In particular, user activity is measured over the same time
horizon, liquidity is expressed in a common numeraire, and the security
and governance indexes are constructed on the same scale. This ensures
that the state variables of the two networks are directly comparable.
The same will be true for the merged blockchain.
In particular, $\sigma_A$, $\sigma_B$ and $\sigma_M$ denote the states of the standalone blockchains and of the merger,
with $\sigma_A=(U_A,L_A,S_A,G_A)$, $\sigma_B=(U_B,L_B,S_B,G_B)$ and $\sigma_M=(U_M,L_M,S_M,G_M)$. We consider non-degenerate merger candidates and therefore assume throughout that $U_i,L_i,S_i,G_i>0$ for $i\in\{A,B\}$.

\subsection{Blockchain valuation function}

We now associate a value with each state $\sigma\in\mathcal D$ through a single (blockchain) valuation function.

\begin{assumption}[Axiomatised valuation]
\label{ass:valuation}
There exists a valuation function\\ $F : \mathcal{D} \to \mathbb{R}^+$ satisfying:
\begin{enumerate}[label=(F\arabic*)]
\item \textbf{Essentiality of users, liquidity, security and governance:}\\ $F(U,L,S,G) = 0$ if and only if $U = 0$, $L = 0$, $S=0$ or $G=0$.\\ A blockchain's value is zero with no users, no liquidity, no security or no governance.
\item \textbf{Monotonicity:} $F$ is non-decreasing in each coordinate.
\item \textbf{Continuity:} $F$ is continuous on $\mathcal{D}$.
\end{enumerate}
\end{assumption}

An immediate consequence of this assumption is that each network has a strictly positive pre-merger value $\mu_A=F(\sigma_A)>0$, $\mu_B=F(\sigma_B)>0$, in accordance with the notation of Section~\ref{sec:mainini}.

(F1)--(F3) deliberately leave the functional form of $F$ otherwise unspecified; calibrating $F$ against empirical blockchain data is future work. These axioms deliberately impose only basic structural restrictions, so any claim below that requires more than (F1)--(F3) is flagged explicitly and not treated as proved.
Indeed, we will see later what additional hypotheses will be needed to prove significant results on the post-merger evaluation.

\subsection{Migration dynamics}

Let $r>0$ be the exchange ratio, in the exact sense of Section~\ref{sec:mainini}: one unit of $A$ becomes one unit of the merged token, one unit of $B$ becomes $r$ units. We introduce migration functions $\tau_A, \tau_B : (0,+\infty) \to [0,1]$, where $\tau_i(r)$ is the fraction of community $i$'s users who migrate to the merged network at exchange rate $r$.

\begin{assumption}[Migration monotonicity and boundary behaviour]
\label{ass:migration}
The migration functions $\tau_A,\tau_B$ are continuous, with $\tau_A$ non-increasing and $\tau_B$ non-decreasing. Moreover, their responses saturate at the extremes:
\[
\lim_{r\to0^+}\tau_A(r)=1,
\quad
\lim_{r\to+\infty}\tau_A(r)=0,\qquad
\lim_{r\to0^+}\tau_B(r)=0,
\quad
\lim_{r\to+\infty}\tau_B(r)=1.
\]
\end{assumption}

\begin{lemma}[Migration-balancing exchange rates]
\label{lem:migrationcrossing}
Under Assumption~\ref{ass:migration}, there exists at least one exchange rate $\tilde r\in(0,+\infty)$ such that $\tau_A(\tilde r)=\tau_B(\tilde r)$. More precisely, the set $Z:=\{r>0:\tau_A(r)=\tau_B(r)\}$ is a nonempty compact interval, and $\tau_A(r)>\tau_B(r)$ for $r<\min Z$, while $\tau_A(r)<\tau_B(r)$ for $r>\max Z$.
\end{lemma}

The boundary conditions have a natural economic interpretation. As
$r\to0^+$, the exchange terms become arbitrarily favourable to
$A$-holders and arbitrarily unfavourable to $B$-holders, leading to full
retention of the former community and full exit of the latter. The
opposite occurs as $r\to+\infty$. Together with continuity and
monotonicity, these limiting responses imply
Lemma~\ref{lem:migrationcrossing}: there is necessarily a nonempty
compact interval of exchange rates at which the two communities exhibit
equal retention.\\
The price-neutral exchange ratio $r_*$
introduced in Section~\ref{sec:mainini} \eqref{eq:price_ratio} 
need not belong to $Z$:
$r_*$ is determined by pre-merger token prices, whereas $Z$ is
determined by the communities' migration responses. To avoid further
complications in this exploratory paper, we impose the following
alignment condition.
\begin{assumption}[Alignment of price and migration neutrality]
\label{ass:alignment}
The exchange ratio $r_*=\frac{p_B}{p_A}$
is also a migration-balancing exchange ratio, i.e.\ $r_*\in Z$. We denote the common retention rate by $\tau^0:=\tau_A(r_*)=\tau_B(r_*)$ and we assume $\tau^0 >0$.
\end{assumption}

The post-merger state is obviously
\begin{equation}
\label{eq:sigmaM}
\sigma_M(r)=\big(U_M(r),L_M(r),S_M(r),G_M(r)\big)
\end{equation}
and we can bound $U_M(r)$ precisely, while for the other three terms we need an 
assumption, as follows in Lemma \ref{lem:Upositive} and Assumption 
\ref{ass:aggregation}.
\begin{lemma}[Bounds on aggregate participation]
\label{lem:Upositive}
For every
$r>0$,
\[
U_A\tau_A(r) + U_B \tau_B (r) \geq U_M(r)
\geq
\max\{U_A\tau_A(r),U_B\tau_B(r)\}
\geq 
\tau^0\min\{U_A,U_B\}
>0,
\]
\end{lemma}

\begin{assumption}[Aggregation]
\label{ass:aggregation}
The post-merger state $\sigma_M(r)$, given by \eqref{eq:sigmaM},
is continuous in $r$ and satisfies $L_M(r)=L_A\tau_A(r)+L_B\tau_B(r)$ and\footnote{The denominators are well defined by Lemma~\ref{lem:Upositive}.}
\[
S_M(r)
=
\frac{
S_AU_A\tau_A(r)+S_BU_B\tau_B(r)
}{U_A\tau_A(r)+U_B\tau_B(r)},
\qquad
G_M(r)
=
\frac{
G_AU_A\tau_A(r)+G_BU_B\tau_B(r)
}{U_A\tau_A(r)+U_B\tau_B(r)
}.
\]
\end{assumption}

Assumption~\ref{ass:aggregation} embodies two modelling
choices. First, the same migration
fraction $\tau_i(r)$ governs the users and liquidity contributed by
community $i$. Second, security and governance are treated as cardinally
comparable across the two networks and aggregate as migrated-user-weighted
means.\footnote{In practice, liquidity may be concentrated among large
holders whose migration decisions differ from those of ordinary users.}.  It can be observed 
that for the normalised security score and for the normalised governance score users of $A$ that are also users of $B$ are considered separately in the 
sum at the numerator and so it is convenient to divide by 
$U_A\tau_A(r)+U_B\tau_B(r)$ instead of the real number of users in $M$.
%

These assumptions give a uniform lower bound on the
post-merger state.
\begin{lemma}[Bounds on aggregate liquidity, security and governance]
\label{lem:Lpositive}
For every
$r>0$, $L_M(r)$, $S_M(r)$ and $G_M(r)$ are positive. More precisely,
$$
L_M(r) \geq\tau^0\min\{L_A,L_B\} \quad
S_M(r) \geq\min\{S_A,S_B\}
\quad
G_M(r) \geq\min\{G_A,G_B\}
$$
\end{lemma}

Since $\tau_A,\tau_B\in[0,1]$ and $S_M(r),G_M(r)$ are
weighted averages,
we can describe the set $\mathcal K$ of admissible states after the merger
$\sigma_M(r)\in\mathcal K:=[0,U_A+U_B]\times[0,L_A+L_B]\times[0,1]^2\subset\mathcal D$ for every $r>0$.
As a consequence, the post-merger state remains in $\mathcal K$, a compact subset of
$\mathcal D$.

\subsection{Endogenous merged value and synergy}
\label{subsec:synergy}

The valuation function $F$ introduced in Assumption~\ref{ass:valuation}
is used throughout to value any network state, whether standalone or
post-merger. Accordingly, the value of the merged network at exchange
rate $r$ is
\[
\mu_M(r)
:=
F\big(\sigma_M(r)\big)
=
F\big(U_M(r),L_M(r),S_M(r),G_M(r)\big).
\]

We define the merger synergy as the difference between the value of the
post-merger network and the combined pre-merger values:
\[
s(r)
:=
\mu_M(r)-\mu_A-\mu_B.
\]
This is formally identical to (\ref{eq:synergy}), but synergy is now endogenous: the exchange rate affects migration,
migration determines the post-merger state, and the post-merger state
determines the merged value.
As a consequence, $s(r)=F(\sigma_M(r))-F(\sigma_A)-F(\sigma_B)$.

Importantly, $s(r)$ has no predetermined sign. If migration leaves the
post-merger network worth less than the two standalone networks
combined, then $s(r)<0$; if the merger creates additional network value,
then $s(r)>0$. 

\begin{proposition}[Continuity and boundedness of merged value and synergy]
\label{prop:sbounded}
Under Assumptions~\ref{ass:valuation}--\ref{ass:aggregation},
the merged value $\mu_M:(0,+\infty)\to\mathbb{R}^+$ is continuous. 
Moreover, let $V_{\max}
:=
\max_{\sigma\in\mathcal K}F(\sigma)
$,
then
\[
0
<
\mu_M(r)
\leq V_{\max}
< 
\infty
\qquad
\text{for every }r>0.
\]
Consequently, $s:(0,\infty)\to\mathbb{R}$ is continuous and satisfies
\[
-(\mu_A+\mu_B)
<
s(r)
\leq
s_{\max}
:=
V_{\max}-(\mu_A+\mu_B)
\qquad
\text{for every }r>0.
\]
\end{proposition}

The proposition guarantees that endogenous synergy remains finite and
varies continuously with the exchange rate. No positivity conclusion is
implied: whether a merger creates or destroys value is an economic
question determined by the post-migration state. The sign of $s(r)$,
and in particular of
$
s_0:=s(r_*),
$
will therefore play a central role
later.

\subsection{Scope of the model}
\label{subsec:scope}

The valuation function $F$ gives the total value of a network from its state and does not depend on token supply, which enters only through the per-token price $p_M(r)=\mu_M(r)/(N_A+rN_B)$: $F$ determines the value of the merged ecosystem, while the exchange ratio and the token supplies determine how that value is allocated across tokens. The migration functions and the state variables $U_M,L_M,S_M,G_M$ are reduced-form descriptors, and no independence among the state coordinates is assumed. The synergy $s(r)=F(\sigma_M(r))-F(\sigma_A)-F(\sigma_B)$ attributes value creation or destruction entirely to the change in network state induced by migration; merger-specific effects not represented by $(U,L,S,G)$, such as integration costs, implementation risk or technological gains, are outside the present framework.

\section{Fairness conditions}
\label{sec:fairness}

From the primitive shareholder-benefit inequalities $p_M \geq p_A$ and $r\,p_M \geq p_B$ of Section~\ref{sec:mainini}, one obtains the 
two inequalities
\begin{equation}
\label{eq:new_barg_ineq}
r\leq r_* \,\frac{\mu_M(r)-\mu_A}{\mu_B},\qquad r(\mu_M(r)-\mu_B)\geq r_*\mu_A.
\end{equation}
If $\mu_M(r)-\mu_B$ is positive, it is possible to obtain the same inequality as in \eqref{eq:over_r_mu} and \eqref{eq:under_r_mu}.  So 
we can consider the following definition, where, if the positivity is not fulfilled at a value $r$, such an exchange rate cannot be in the 
wealth preservation set.

\begin{definition}[Wealth preservation]
\begin{align*}
\mathcal{R}_{\mathrm{wealth}} &:= \{\, r>0 \;:\; r \in \mathcal{BR}_\mu(\mu_M(r)) \,\}\\
&\phantom{:}= \{\, r>0 \;:\; \underline{r}_\mu(\mu_M(r)) \leq r \leq \overline{r}_\mu(\mu_M(r)) \,\},
\end{align*}
using $\mathcal{BR}_\mu$, $\underline{r}_\mu$, $\overline{r}_\mu$ exactly as recalled in Section~\ref{sec:mainini}, and $\mu_M(r)$ as constructed in Section~\ref{sec:setup}.
\end{definition}

\begin{definition}[Adoption preservation]
\label{def:adoption}
Fix a retention threshold $\tau^\ast \in (0,1)$:
\[
\mathcal{R}_{\mathrm{adoption}} := \{\, r>0 : \tau_A(r) \geq \tau^\ast \text{ and } \tau_B(r) \geq \tau^\ast \,\}.
\]
\end{definition}

\begin{definition}[Security preservation]
Fix a critical security level $S^\ast \in (0,1)$:
\[\mathcal{R}_{\mathrm{security}} := \{\, r>0 : S_M(r) \geq S^\ast \,\}.\]
\end{definition}

\begin{definition}[Governance preservation]
\label{def:governance}
Fix a critical governance-\hspace{0pt}participation level $G^\ast \in (0,1)$:
\[\mathcal{R}_{\mathrm{governance}} := \{\, r>0 : G_M(r) \geq G^\ast \,\}.\]
\end{definition}

Definition~\ref{def:governance} is a threshold condition on the post-merger governance-participation score, parallel to the security condition, and it is the reading consistent with the state variable actually defined in Section~\ref{sec:setup}: $G$ was introduced as a normalised measure of \emph{participation}, not of the \emph{distribution} of voting power between the two communities.

\begin{definition}[Liquidity preservation]
Fix a minimal post-merger liquidity $L^\ast>0$:
\[\mathcal{R}_{\mathrm{liquidity}} := \{\, r>0 : L_M(r) \geq L^\ast \,\}.\]
\end{definition}

\begin{definition}[Admissible exchange rates]
\[
\mathcal{R} := \mathcal{R}_{\mathrm{wealth}} \cap \mathcal{R}_{\mathrm{adoption}} \cap \mathcal{R}_{\mathrm{security}} \cap \mathcal{R}_{\mathrm{governance}} \cap \mathcal{R}_{\mathrm{liquidity}}.
\]
\end{definition}

\subsection{Structure of the non-wealth conditions}
\label{sec:static}

\begin{proposition}[Interval structure and endpoints]
\label{prop:intervals}
Under Assumptions~\ref{ass:valuation}--\ref{ass:aggregation}:
\begin{enumerate}[label=(\alph*)]
\item $\mathcal{R}_{\mathrm{adoption}} = [\underline r, \overline r]$ for some $0 < \underline r \leq \overline r < +\infty$ or is empty; in particular it is \emph{compact}.
\item $S_M(\cdot)$ and $G_M(\cdot)$ are continuous and monotone on $(0,+\infty)$, with
\[
\lim_{r\to0^+}\big(S_M,G_M\big)(r) = (S_A,G_A), \qquad \lim_{r\to+\infty}\big(S_M,G_M\big)(r) = (S_B,G_B),
\]
their ranges are contained in the intervals $[\min\{S_A,S_B\},\max\{S_A,S_B\}]$ and $[\min\{G_A,G_B\},$ $\max\{G_A,G_B\}]$ respectively. Consequently, 
$\mathcal{R}_{\mathrm{security}}$ (respectively, $\mathcal{R}_{\mathrm{governance}}$)  is either a closed interval or all of $(0,+\infty)$, which happens only when $S^* \leq 
\min\{S_A,S_B\}$ (respectively $G^* \leq \min\{G_A,G_B\}$), or the empty set, which happens only when $S^*\geq\max\{S_A,S_B\}$ (respectively $G^* \geq \max\{G_A,G_B\}$).
\item $L_M(\cdot)$ is continuous with $\lim_{r\to0^+}L_M(r) = L_A$, $\lim_{r\to+\infty}L_M(r) = L_B$, $L_M(r_*) = \tau^0(L_A+L_B)$ and $\inf_{r>0}L(r) \geq \tau^0
\min\{L_A,L_B\} > 0$. It need 
\emph{not} be monotone, and $\mathcal{R}_{\mathrm{liquidity}}$ need not be an interval; it is all of $(0,+\infty)$ whenever $L^\ast \leq \tau^0\min\{L_A,L_B\}$.
\end{enumerate}
\end{proposition}

Part (c) is worth emphasizing, since it identifies liquidity as the one channel through which the admissible set can genuinely disconnect: the other three non-wealth conditions
are monotone in $r$ and therefore contribute intervals, whereas $L_M$ is a sum of a non-increasing and a non-decreasing function and is hump- or dip-shaped according to the size of 
$\tau^0$. Non-monotonicity is genuine, not a gap in the argument: taking $\tau^0$ close to $1$ gives $L_M(r_*) \approx L_A + L_B > \max\{L_A,L_B\} = \max\{\lim_{r\to0^+}
L_M(r),\lim_{r\to+\infty}L_M(r)\}$, so $L_M$ rises and then falls; taking $\tau^0$ small gives $L_M(r_*) =\tau^0(L_A+L_B) < \min\{L_A,L_B\}$, so $L_M$ falls and then rises, and for $L^\ast$ 
between $\tau^0(L_A+L_B)$ and $\min\{L_A,L_B\}$ the set $\{L_M \geq L^\ast\}$ has two connected components.
The economic content of the dip-shaped case is that a mediocre retention rate can leave the merged pool thinner than either original pool, even though the extreme rates preserve one pool intact.

\subsection{The self-referential wealth condition}
\label{subsec:selfref}

The definition of $\mathcal{R}_{\mathrm{wealth}}$ from the primitive shareholder-benefit inequalities gives \eqref{eq:new_barg_ineq}
and using $\mu_M(r)-\mu_B = \mu_A+s(r)$ and $\mu_M(r)-\mu_A = \mu_B+s(r)$, one obtains
\begin{equation}
\label{eq:primitive}
r\,\big(\mu_A + s(r)\big) \;\geq\; r_*\,\mu_A
\qquad\text{and}\qquad
r \;\leq\; r_*\Big(1+\frac{s(r)}{\mu_B}\Big).
\end{equation}
Since $s(r)$ may now be negative, some care is needed: whenever $\mu_A + s(r) > 0$, the first inequality is equivalent to the familiar ratio form $r \geq r_*\mu_A/(\mu_A+s(r))$; 
whenever $\mu_A + s(r) \leq 0$ (merged value not exceeding $\mu_B$ alone), the first inequality fails outright, so such $r$ never belong to $\mathcal{R}_{\mathrm{wealth}}$. 
The bound becomes
\begin{equation}
\label{eq:wealthband}
r_*\,\frac{\mu_A}{\mu_A + s(r)} \;\leq\; r \;\leq\; r_*\Big(1 + \frac{s(r)}{\mu_B}\Big).
\end{equation}
and is self-referential in $r$ through $s(r)$. We treat each part of (\ref{eq:wealthband}) as a fixed-point equation.

\begin{definition}
\label{def:phi}
On $\{r > 0 : \mu_A + s(r) > 0\}$, set
\[
\Phi_B(r) := r_*\cdot\frac{\mu_A}{\mu_A+s(r)}, \qquad
\Phi_A(r) := r_*\Big(1+\frac{s(r)}{\mu_B}\Big),
\]
so that, on that set, membership in $\mathcal{R}_{\mathrm{wealth}}$ reads $\Phi_B(r) \leq r \leq \Phi_A(r)$.
\end{definition}

The subscript names the community whose wealth-preservation requirement generates the bound, not the parameter appearing in it: $A$-shareholders' condition $p_M \geq p_A$ caps $r$ 
from above, giving the upper bound $\Phi_A$, which involves $\mu_B$; $B$-shareholders' condition $r\,p_M \geq p_B$ bounds $r$ from below, giving $\Phi_B$, which involves $\mu_A$.

\begin{proposition}[Existence of self-consistent rates]
\label{prop:selfref}
Suppose, in addition to Assumptions~\ref{ass:valuation}--\ref{ass:aggregation}, the no-catastrophe bound
\[
s_{\min} := \inf_{r>0} s(r) \;>\; -\min\{\mu_A, \mu_B\},
\]
i.e.\ the merged value always exceeds the larger of the two standalone values. Then there exist $\underline{r}^{\,\ast}, \overline{r}^{\,\ast} \in (0,+\infty)$ with
\[
\underline{r}^{\,\ast} = \Phi_B(\underline{r}^{\,\ast}), \qquad \overline{r}^{\,\ast} = \Phi_A(\overline{r}^{\,\ast}),
\]
i.e.\ exchange rates that saturate, respectively, the $B$-side and $A$-side wealth-preservation bounds self-consistently against the migration they themselves induce.
\end{proposition}

Unlike a generic fixed-point map postulated to have an invariant compact interval, $\Phi_A$ and $\Phi_B$ are built explicitly from $r_*, \mu_A, \mu_B$ and $s(r)$, and the invariant intervals are derived from the boundedness established in Proposition~\ref{prop:sbounded}; but the no-catastrophe bound is a genuine hypothesis — it fails precisely for mergers whose worst-case migration collapse destroys most of the combined value.

If $s(r) \geq 0$ for all $r$ (the merger is value-creating at every rate), then $\Phi_A(r) \geq r_*$ and $\Phi_B(r) \leq r_*$ pointwise, so the two saturating rates bracket the pre-merger price ratio; for sign-changing $s$ no such bracketing is claimed. Whether — and how widely — the two bounds of \eqref{eq:primitive} are satisfied \emph{simultaneously}, i.e.\ the structure of $\mathcal{R}_{\mathrm{wealth}}$ itself, is the main problem of the paper. It is resolved in the next section, where we show that concrete valuation functions proposed in the literature share a structural property strong enough to settle it.

In the following proposition an important result is proved on the exchange rates in the bargaining region.  Also the question whether the unbiased rate 
$r_*$ is in $\mathcal{R}_{\mathrm{wealth}}$ is answered.

\begin{proposition}[Value-creation dichotomy]
\label{prp:value_creation}
    Let Assumptions~\ref{ass:migration}--\ref{ass:aggregation} (with $\tau^0 > 0$) hold, together with the standing positivity of the pre-merger state. Then 
    $\mathcal{R}_{\mathrm{wealth}} \subseteq \{r > 0 : s(r) \geq 0\}$. Moreover $r_* \in \mathcal{R}_{\mathrm{wealth}}$ if and only if $s_0 \geq 0$.
\end{proposition}

In particular no exchange rate at which the merger destroys value is wealth-admissible.  The price ratio is wealth-admissible if and only if the merger is value-creating at the unbiased rate.

\section{A common point: the unbiased rate is admissible}
\label{sec:r*}

The boundary and anchoring structure allows to exhibit a single rate belonging to all 
five sets. The natural candidate is $r_*$, and the conditions under which it works 
are exactly that each community's threshold be one the merged network actually 
delivers at the neutral rate.

\begin{definition}[Threshold admissibility at $r_*$]
\label{def:thresh}
The threshold profile $(\tau^\ast, S^\ast,\allowbreak G^\ast, L^\ast)$ is \emph{admissible at $r_*$} if
\[
\tau^\ast \leq \tau^0, \qquad
S^\ast \leq S_M(r_*), \qquad
G^\ast \leq G_M(r_*),
\]
\[
L^\ast \leq \tau^0(L_A+L_B),
\]
and \emph{strictly admissible} if all four inequalities are strict.
\end{definition}

Note that $S_M(r_*) = (S_AU_A+S_BU_B)/(U_A+U_B)$ and $G_M(r_*) = (G_AU_A+G_BU_B)/(U_A+U_B)$ are the user-weighted means of the pre-merger scores, computable from observables before any merger takes place; likewise $\tau^0$ and $\tau^0(L_A+L_B)$. Threshold admissibility is therefore a checkable condition on the merger candidates, not an abstract hypothesis.

\begin{theorem}[Nonemptiness of the full admissible set]
\label{thm:common}
Let Assumptions~\ref{ass:valuation}--\ref{ass:aggregation} hold, let the threshold profile be admissible at $r_*$, and let $s_0 \geq 0$. Then
\[
r_* \in \mathcal{R} = \mathcal{R}_{\mathrm{wealth}} \cap \mathcal{R}_{\mathrm{adoption}} \cap \mathcal{R}_{\mathrm{security}} \cap \mathcal{R}_{\mathrm{governance}} \cap \mathcal{R}_{\mathrm{liquidity}};
\]
in particular $\mathcal{R} \neq \emptyset$.
\end{theorem}

Theorem~\ref{thm:common} is the existence statement the framework has been aiming at since Section~\ref{sec:fairness}: not merely that the wealth band is nonempty, but that all five fairness requirements can be met simultaneously, at an explicitly identified rate. 
Its hypotheses also make the failure modes precise. The neutral rate is not fair when either (i) the merger destroys value at that rate ($s_0 < 0$), or (ii) some community demands a threshold that the merged network cannot deliver at $r_*$ — a demand for more security, governance participation, liquidity or retention than the user-weighted combination of the two predecessors provides. Neither failure is a defect of the model; both are the model correctly reporting that the merger should not proceed at $r_*$.

\section{Literature proposals for $F$, a shared structural property, and the Main Theorem}
\label{sec:mainthm}

In this section we consider functions that belong to the following class.
\begin{definition}[Network-convex class]
\label{def:classC}
Let $\mathcal{C}$ be the set of functions $f : [0,+\infty) \to [0,+\infty)$ that are continuous, nondecreasing, convex, satisfy $f(0)=0$, and are strictly positive on $(0,
+\infty)$.
\end{definition}

\subsection{Four proposals from the literature}
\label{subsec:proposals}

The network-economics and cryptocurrency literatures propose four principal laws for the value of a network as a function of its user count $U$:
\begin{enumerate}[label=(P\arabic*)]
\item the \textbf{Sarnoff-type linear law} $f(U)=cU$, the lower benchmark in \cite{BriscoeOdlyzkoTilly2006};
\item the \textbf{Odlyzko--Tilly law} $f(U)=cU\log U$ \cite{BriscoeOdlyzkoTilly2006}, a Zipf-motivated correction to Metcalfe's law, used here in the normalised form $f(U)=cU\log(1+U)$, which is finite, nonnegative and vanishes at $0$;
\item \textbf{Metcalfe's law} $f(U)=cU^2$ \cite{Metcalfe2013}, applied to Bitcoin in \cite{Peterson2018} and to several blockchain networks in \cite{Alabi2017};
\item the \textbf{generalised Metcalfe law} $f(U)=cU^\beta$ with $\beta\in[1,2]$, where $\beta\approx1.69$ is estimated for Bitcoin in \cite{Wheatley2019}.
\end{enumerate}
These are laws for the \emph{user-network component} of value: $cU^\beta$ is positive whenever $U>0$, even if $L=0$, $S=0$ or $G=0$, so it violates (F1) and is not a valuation function. Accordingly, (P1)--(P4) are candidates for the factor $f$ in the following assumption.

\begin{assumption}[Structured valuation]
\label{ass:structured}
We assume that
$$F(U,L,S,G) = f(U)\,\psi(L,S,G)\,,$$ 
where:
\begin{enumerate}[label=(\roman*)]
\item $f$ belongs to the \emph{network-convex class} $\mathcal{C}$ defined in Definition \ref{def:classC};
\item $\psi : \mathbb{R}_{\geq 0}\times[0,1]^2 \to [0,1]$ is continuous, nondecreasing in each argument, and $\psi(L,S,G) = 0$ if and only if $L = 0$, $S=0$ or $G=0$.
\end{enumerate}
\end{assumption}

Any such $F$ satisfies (F1)--(F3): since $f(0)=0$ and $f>0$ on $(0,+\infty)$, $f$ handles the boundary $U=0$ and $\psi$ the others, while monotonicity and continuity are inherited factorwise. The proposals (P1)--(P4) correspond to $\psi\approx1$; the results below use only $\psi\in[0,1]$ and the boundary characterisation in (ii).

\subsection{The shared structural property}

Each of (P1)--(P4) belongs to $\mathcal{C}$: linearity is (weak) convexity; $U\log(1+U)$ has second derivative $c[(1+U)^{-1} + (1+U)^{-2}] > 0$; and $U^\beta$ is convex for every $\beta \geq 1$. The class is thus exactly the common structural core of the four proposals, and the following elementary lemma extracts everything we use from it.

\begin{lemma}[Consequences of network-convexity]
\label{lem:classC}
Let $f \in \mathcal{C}$. Then:
\begin{enumerate}[label=(\alph*)]
\item (Sublinearity toward the origin) $f(\lambda x) \leq \lambda f(x)$ for all $\lambda \in [0,1]$, $x \geq 0$.\\
If $f$ is strictly convex then $f(\lambda x) < \lambda f(x)$ for all $\lambda \in (0,1)$, $x > 0$.
\item (Superadditivity) $f(x+y) \geq f(x) + f(y)$ for all $x, y \geq 0$, with strict inequality for $x,y > 0$ whenever $f$ is strictly convex.
\item (Automatic Lipschitz bounds on compacts) For every $M > 0$, $f$ is Lipschitz on $[0,M]$ with constant $\Lambda_f(M) := f(M+1) - f(M)$.
\end{enumerate}
\end{lemma}

\subsection{The Main Theorem}

\begin{theorem}[Main Theorem]
\label{thm:main}
Let Assumptions~\ref{ass:migration}--\ref{ass:aggregation} (with $\tau^0 > 0$) and~\ref{ass:structured} hold, together with the standing positivity of the pre-merger state. Then:
\begin{enumerate}[label=(\roman*)]
\item \textbf{(Nondegeneracy of $\mathcal{R}_{\mathrm{wealth}}$)} If $s_0 > 0$, there exist $r^- < r_* < r^+$ with $[r^-, r^+] \subseteq \mathcal{R}_{\mathrm{wealth}}$.
\item \textbf{(Quantitative width for $\mathcal{R}_{\mathrm{wealth}}$)} If $s_0 > 0$ and moreover $\tau_A, \tau_B$ and $\psi$ are Lipschitz, then $s$ is Lipschitz on $(0, +\infty)$ 
with some constant $\Lambda\in\mathbb{R}$ determined by the primitives, and
\[
\big[\,r_* - \varepsilon_-,\; r_* + \varepsilon_+\,\big] \subseteq \mathcal{R}_{\mathrm{wealth}},
\]
where
\[
\varepsilon_+ = \frac{r_* s_0}{\mu_B + r_*\Lambda},
\qquad
\varepsilon_- = \frac{r_* s_0}{\mu_A + s_0 + r_*\Lambda}.
\]
\item \textbf{(Nondegeneracy of $\mathcal{R}$)} If the profile is strictly admissible and $s_0 > 0$, then $\mathcal{R}$ contains a nondegenerate closed interval around $r_*$: 
there is $\varepsilon>0$ with $[r_*-\varepsilon, r_*+\varepsilon] \subseteq \mathcal{R}$, and $\varepsilon$ may be taken to be the minimum of the five individual half-widths, the 
wealth one being given explicitly in (ii).
\end{enumerate}
\end{theorem}

\begin{remark}
Proposition \ref{prp:value_creation} recovers, endogenously, exactly the dichotomy of \cite{Mainini2024}: their bargaining region $\mathcal{BR}_\mu(\mu_M)$ is nonempty precisely when $\mu_M \geq \mu_A + \mu_B$, and our result says the same with $\mu_M$ now determined by the migration that $r$ itself induces. Value-destroying mergers admit no fair exchange rate and parts (ii)--(iii) quantify how strictly positive synergy at $r_*$ fattens the admissible set into a genuine interval around it, of width controlled explicitly by $s_0$, $\mu_A$, $\mu_B$ and the migration sensitivity $\Lambda$.
\end{remark}

\section{Conclusions and open problems}
\label{sec:conclusions}

We have extended the exchange-ratio framework of \cite{Mainini2024} to cryptocurrency mergers by making the merged value endogenous: the exchange ratio $r$ determines migration, migration determines (up to user overlap) the post-merger state $\sigma_M(r)$, and the post-merger value $\mu_M(r)=F(\sigma_M(r))$. The synergy $s(r)=\mu_M(r)-\mu_A-\mu_B$ has no predetermined sign, and the bounds of the wealth condition become self-referential in $r$.

No exchange rate at which the merger destroys value is wealth-admissible, and the price ratio $r_*$ is wealth-admissible \emph{exactly when} $s_0\geq0$ (Proposition~\ref{prp:value_creation}). Under reasonable assumptions, $r_*$ satisfies all five fairness conditions simultaneously (Theorem~\ref{thm:common}); under strict admissibility and $s_0>0$, the admissible set contains a nondegenerate interval around $r_*$, with wealth half-widths explicit in $s_0$, $\mu_A$, $\mu_B$ and $\Lambda$ (Theorem~\ref{thm:main}). Liquidity is the one (non-wealth) channel through which the admissible set can disconnect: a low retention rate can leave the merged pool thinner.

Several problems remain open.

\emph{Value creation.} Superadditivity of $f\in\mathcal{C}$ suggests a critical retention rate above which $s_0\geq0$; deriving it requires control of user overlap, since $U_M(r)$ may lie anywhere between $\max\{U_A\tau_A(r),U_B\tau_B(r)\}$ and $U_A\tau_A(r)+U_B\tau_B(r)$, and superadditivity gives no gain when the two communities largely coincide.

\emph{Calibration and estimation.} The valuation function must reproduce the observed capitalisations $\mu_A,\mu_B$, which requires order consistency (if $\sigma_A$ dominates $\sigma_B$ coordinatewise then $\mu_A\geq\mu_B$), a range condition on the chosen family and, since a single merger traces only a curve in $\mathcal{D}$, cross-sectional data. The migration functions could be estimated from historical token migrations and contested forks \cite{BiaisEtAl2019}, pooled in the normalised rate $r/r_*$; this would pin down $\tau^0$ and (combined with $F$) $\Lambda$, and show whether liquidity thresholds in the window $\big(\tau^0(L_A+L_B),\,\min\{L_A,L_B\}\big)$, where the liquidity set splits, occur in practice.

\emph{Microfoundations.} Deriving $\tau_A,\tau_B$ from a participation game with network externalities \cite{SockinXiong2023,CongLiWang2021} would test Assumptions~\ref{ass:migration} and~\ref{ass:alignment}; coordination-driven collapse of participation would break continuity and could disconnect the admissible set.

\emph{Scope.} Parity of governance weight between the two communities, merger costs and benefits not captured by the state $(U,L,S,G)$, and risk-related variables are natural extensions.

\section*{Acknowledgments}

Anthropic Claude and OpenAI ChatGPT reviewed the arguments and assisted with some proofs.

\appendix
\section{Proofs}
\label{app:proofs}

\begin{proofof}{Lemma~\ref{lem:migrationcrossing}}
Define
$
d(r):=\tau_A(r)-\tau_B(r)
$.\\
By Assumption~\ref{ass:migration}, $d$ is continuous and non-increasing on
$(0,+\infty)$. Moreover,
\[
\lim_{r\to0^+}d(r)=1,
\qquad
\lim_{r\to+\infty}d(r)=-1.
\]
Hence $d(r)=0$ for at least
one $r\in(0,+\infty)$, so $Z$ is nonempty.

Since $d$ is continuous, $Z$ is closed relative to $(0,+\infty)$.
The limiting signs imply that $Z$ is bounded away from both $0$ and
$+\infty$, and hence $Z$ is compact in $\mathbb{R}$. Since $d$ is
non-increasing, if $r_1,r_2\in Z$ with $r_1<r_2$, then
\[
0=d(r_1)\ge d(r)\ge d(r_2)=0
\]
for every $r\in[r_1,r_2]$. Thus $d(r)=0$ in $[r_1,r_2]$ and
$Z$ is an interval. Outside $Z$ these inequalities  follow
immediately from monotonicity and the limiting signs.
\end{proofof}

\begin{proofof}{Lemma~\ref{lem:Upositive}}
If all users of $A$ that move to the new cryptocurrency are not users of $B$, then obviously $U_M(r)=U_A\tau_A(r)+U_B\tau_B(r)$.
Otherwise, they are fewer, but certainly at least the number of those of $A$ moving into $M$, i.e. $U_A\tau_A(r)$, and similarly for $B$, i.e $U_B\tau_B(r)$, so 
$$ U_A\tau_A(r) + U_B \tau_B (r) \geq U_M(r) \geq \max\{U_A\tau_A(r),U_B\tau_B(r)\}$$

Suppose first that $r\leq r_*$. Since $\tau_A$ is non-increasing,
$\tau_A(r_*)=\tau^0>0$ and $U_A>0$, we have
\[
\tau_A(r)\geq\tau_A(r_*)=\tau^0
>0\qquad
U_A\tau_A(r) \geq U_A \tau^0>0
\]

If instead $r\geq r_*$, we arrive similarly at
$\tau_B(r_*)=\tau^0$, we have
\[
\tau_B(r)\geq\tau_B(r_*)=\tau^0
\qquad
U_B\tau_B(r) \geq U_B \tau^0 >0
\]
In either case,
$\max\{\tau_A U_A,\tau_B U_B\}
\geq 
\tau^0\min\{U_A,U_B\}
>0$.
\end{proofof}

\begin{proofof}{Lemma~\ref{lem:Lpositive}}
As in Lemma~\ref{lem:Upositive},
if $r\leq r_*$ then
$\tau_A(r)\geq \tau^0$,
if instead $r\geq r_*$, then 
$\tau_B(r)\geq\tau^0$.
In either case, from 
$\tau^0>0$ we get
\[
L_M(r)\geq\tau^0\min\{L_A,L_B\}>0\,.
\]
Now, let $S'=\min\{S_A,S_B\}$
$$
S_M(r)
=
\frac{
S_AU_A\tau_A(r)+S_BU_B\tau_B(r)
}{
U_A\tau_A(r)+U_B\tau_B(r)
}\geq 
\frac{
S'\left(U_A\tau_A(r)+U_B\tau_B(r)\right)
}{
U_A\tau_A(r)+U_B\tau_B(r)
}
=S'
$$
Similarly we can argue for $G_M(r)$.
\end{proofof}

\begin{proofof}{Proposition~\ref{prop:sbounded}}
By Assumption~\ref{ass:aggregation} and
Lemma~\ref{lem:Upositive}, the post-merger state map
$
r\longmapsto\sigma_M(r)
$
is continuous on $(0,+\infty)$. Since $F$ is continuous,
$
\mu_M(r)=F\big(\sigma_M(r)\big)
$
is so.

Moreover, the image of the post-merger state map is contained in the
compact set $\mathcal K$ defined above. 
By Lemmas \ref{lem:Upositive} and \ref{lem:Lpositive}, all four coordinates of $\sigma_M(r)$ are strictly positive, so the
continuity of $F$ 
implies, by the extreme value theorem,
\[
0
<
F\big(\sigma_M(r)\big)
\leq
\max_{\sigma\in\mathcal K}F(\sigma)
=
V_{\max}
<
\infty.
\]
Finally, since
$
s(r)=\mu_M(r)-\mu_A-\mu_B,
$
continuity of $s$ and the stated bounds follow immediately.
\end{proofof}

\begin{proofof}{Proposition~\ref{prop:intervals}}
(a) Being $\tau_B$ continuous and non-decreasing gives that 
$\{r:\tau_B(r)\geq\tau^\ast\}$ is a 
closed half-line $[\underline r,+\infty)$ and $\underline r > 0$ because 
$\lim_{r\to0^+}\tau_B(r) = 0 < \tau^\ast$. Being $\tau_A$ continuous and 
non-increasing 
gives $\{r:\tau_A(r)\geq\tau^\ast\} = (0,\overline r\,]$ and $\overline 
r\in\mathbb{R}$, because $\lim_{r\to+\infty}\tau_A(r) = 0 < \tau^\ast$. The 
intersection is $[\underline r, \overline r]$ or empty if $\underline r > \overline 
r$.  It is a compact set.

(b) Write $a(r) := U_A\tau_A(r)$ and $b(r) := U_B\tau_B(r)$, so that by Assumption~\ref{ass:aggregation}
\[
S_M(r) = S_A + (S_B - S_A)\,w(r), \qquad w(r) := \frac{b(r)}{a(r)+b(r)},
\]
with $a+b > 0$ everywhere by Lemma~\ref{lem:Upositive}. We show $w$ is non-decreasing without dividing by possibly-vanishing quantities. Let $r_1 < r_2$. Then $b(r_1) \leq b(r_2)$ and $a(r_2) \leq a(r_1)$, hence
\[
b(r_1)a(r_2) \;\leq\; b(r_2)a(r_1).
\]
Adding $b(r_1)b(r_2)$ to both sides and factoring gives $b(r_1)\big(a(r_2)+b(r_2)\big) \leq b(r_2)\big(a(r_1)+b(r_1)\big)$; dividing by the strictly positive product $\big(a(r_1)+b(r_1)\big)\big(a(r_2)+b(r_2)\big)$ yields $w(r_1) \leq w(r_2)$. Thus $w$ is non-decreasing, and $S_M$ is monotone — non-decreasing if $S_B > S_A$, non-increasing if $S_B < S_A$, constant if $S_A = S_B$. 
Function $w(r)$ is continuous, since $w$ is a quotient of continuous functions with denominator bounded away from $0$ by Lemma~\ref{lem:Upositive}. The boundary limits follow from 
Assumption~\ref{ass:migration}: as $r\to0^+$, $b\to0$ and $a\to U_A>0$, so $w\to0$ and $S_M\to S_A$; as $r\to\infty$, $a\to0$ and $b\to U_B>0$, so $w\to1$ and $S_M\to S_B$. 
Since $w\in[0,1]$, $S_M(r)$ is a convex combination of $S_A,S_B$, giving the range bound. A superlevel set of a continuous monotone function is a closed half-line (or empty, or 
everything), hence an interval; and if $S^\ast \leq \min\{S_A,S_B\}$ then $S_M(r)\geq\min\{S_A,S_B\}\geq S^\ast$ for all $r$. A similar situation holds for the empty set.  The 
argument for $G_M$ is identical with $(G_A,G_B)$ in place of $(S_A,S_B)$.

(c) Continuity derives from the assumptions and the lower bound comes from Lemma~\ref{lem:Lpositive}; the limits and the value at $r_*$ are immediate from 
Assumptions~\ref{ass:migration} and \ref{ass:alignment}.  The final claim is the lower bound.
\end{proofof}

\begin{proofof}{Proposition~\ref{prop:selfref}}
By Proposition~\ref{prop:sbounded}, $s$ is continuous with $s_{\min} \leq s(r) \leq s_{\max}$ for every $r>0$, and by hypothesis $\mu_B + s_{\min} > 0$ and $\mu_A + s_{\min} > 0$, 
so both $\Phi_A$ and $\Phi_B$ are defined and continuous on all of $(0,+\infty)$. The range of $\Phi_A$ is contained in
\[
[a,b] := \Big[r_*\big(1+\tfrac{s_{\min}}{\mu_B}\big),\; r_*\big(1+\tfrac{s_{\max}}{\mu_B}\big)\Big] \subset (0,+\infty),
\]
with $a > 0$ by the no-catastrophe bound; in particular $\Phi_A$ maps the compact interval $[a,b]$ into itself. Let $g(r) := \Phi_A(r)-r$ on $[a,b]$. Since $\Phi_A(a) \geq a$ (as $a$ is the infimum of $\Phi_A$'s range) and $\Phi_A(b) \leq b$ (as $b$ is the supremum of that range), $g(a)\geq 0$ and $g(b)\leq 0$. Since $g$ is continuous, the intermediate value theorem gives $\overline{r}^{\,\ast}\in[a,b]$ with $g(\overline{r}^{\,\ast})=0$.

For $\Phi_B$: it is continuous with range contained in $[a',b'] := \big[r_*\mu_A/(\mu_A+s_{\max}),\, r_*\mu_A/(\mu_A+s_{\min})\big] \subset (0,+\infty)$, both endpoints 
positive and finite by the no-catastrophe bound. The identical infimum/supremum argument and the intermediate value theorem give $\underline{r}^{\,\ast}\in[a',b']$ with 
$\Phi_B(\underline{r}^{\,\ast})=\underline{r}^{\,\ast}$.
\end{proofof}

\begin{proofof}{Proposition~\ref{prp:value_creation}}
Let $r \in \mathcal{R}_{\mathrm{wealth}}$. By the discussion following \eqref{eq:primitive}, $\mu_A + s(r) > 0$ and the bound \eqref{eq:wealthband} holds at $r$. If $s(r) < 0$, then $\mu_A/(\mu_A + s(r)) > 1$ and $1 + s(r)/\mu_B < 1$, so the band gives $r > r_*$ and $r < r_*$ simultaneously — a contradiction; hence $s(r) \geq 0$. For the second claim: if $s_0 \geq 0$ then $\mu_A/(\mu_A + s_0) \leq 1$ and $1 + s_0/\mu_B \geq 1$, so both inequalities of \eqref{eq:wealthband} hold at $r = r_*$, i.e.\ $r_* \in \mathcal{R}_{\mathrm{wealth}}$; the converse is the inclusion just proved, applied at $r = r_*$.
\end{proofof}

\begin{proofof}{Theorem~\ref{thm:common}}
Membership of $r_*$ in each set: $\tau_A(r_*) = \tau_B(r_*) = \tau^0 \geq \tau^\ast$ gives $r_* \in \mathcal{R}_{\mathrm{adoption}}$; $S_M(r_*) \geq S^\ast$ and $G_M(r_*) \geq G^\ast$ are the second and third admissibility inequalities; $L(r_*) = \tau^0(L_A+L_B) \geq L^\ast$ is the fourth; and $s_0 \geq 0$ gives $r_* \in \mathcal{R}_{\mathrm{wealth}}$ by 
Proposition~\ref{prp:value_creation}. Hence $r_*$ lies in the intersection.
\end{proofof}

\begin{proofof}{Lemma~\ref{lem:classC}}
\textbf{(a)} By convexity and $f(0)=0$, for every
$\lambda\in[0,1]$ and $x\geq0$,
\[
f(\lambda x)
=
f\bigl(\lambda x+(1-\lambda)0\bigr)
\leq
\lambda f(x)+(1-\lambda)f(0)
=
\lambda f(x).
\]
If $f$ is strictly convex, $x>0$, and $\lambda\in(0,1)$, then
$x$ and $0$ are distinct and so
\[
f(\lambda x)=f\bigl(\lambda x+(1-\lambda)0\bigr)
<
\lambda f(x)+(1-\lambda)f(0)
=
\lambda f(x) \quad
\implies f(\lambda x)<\lambda f(x)\,.
\]
\textbf{(b)} If $x+y=0$, so $x=y=0$, the claim is trivial. Otherwise, note that
\[
\lambda:=\frac{x}{x+y}\in[0,1]
\implies \qquad
x=\lambda(x+y),
\quad
y=(1-\lambda)(x+y).
\]
By part~(a),
\[
f(x)
=
f\bigl(\lambda(x+y)\bigr)
\leq
\lambda f(x+y)
\qquad
f(y)
=
f\bigl((1-\lambda)(x+y)\bigr)
\leq
(1-\lambda)f(x+y)\,.
\]
Summing the two inequalities gives
\[
f(x)+f(y)\leq f(x+y).
\]
If $f$ is strictly convex and $x,y>0$, then
$\lambda\in(0,1)$ and $1-\lambda\in(0,1)$. Hence the strict part of
(a), applied with $x+y>0$, gives the strict version of the previous inequality.

\textbf{(c)} For $0\leq x<y\leq M$, monotonicity of $f$ and the
three-chord inequality for convex functions give
\[
0
\leq
\frac{f(y)-f(x)}{y-x}
\leq
\frac{f(M)-f(y)}{M-y}
\leq
\frac{f(M+1)-f(M)}{(M+1)-M}
=
\Lambda_f(M).
\]
Hence
\[
|f(y)-f(x)|
\leq
\Lambda_f(M)\,|y-x|,
\]
so $f$ is Lipschitz on $[0,M]$ with Lipschitz constant
$\Lambda_f(M)=f(M+1)-f(M)$.
\end{proofof}

\begin{proofof}{Theorem~\ref{thm:main}}
(i)  If $s_0 > 0$, both inequalities in \eqref{eq:wealthband} are \emph{strict} at $r = r_*$. The functions $g_1(r) := r - r_*\mu_A/(\mu_A+s(r))$ and $g_2(r) := r_*(1+s(r)/\mu_B) - r$ are continuous on the open set $\{\mu_A + s > 0\} \ni r_*$ (continuity of $s$: Proposition~\ref{prop:sbounded}) and strictly positive at $r_*$; hence both remain positive on some closed interval $[r_*-\varepsilon, r_*+\varepsilon]$ contained in that set, with $\varepsilon > 0$, which is then inside $\mathcal{R}_{\mathrm{wealth}}$.

(ii) \emph{Step 1 ($s$ is Lipschitz on $(0,+\infty)$).} By Lemma~\ref{lem:Upositive}, $U_M(r) \geq \tau^0\min\{U_A,U_B\} > 0$ for every $r > 0$. The maps $U_M(\cdot), L_M(\cdot)$ are Lipschitz on $(0,+\infty)$ (finite nonnegative combinations of the Lipschitz $\tau_A,\tau_B$); $S_M(\cdot), G_M(\cdot)$ are quotients of bounded Lipschitz functions whose denominator is bounded below by Lemma \ref{lem:Lpositive}, hence Lipschitz on $(0,+\infty)$; $f$ is Lipschitz on the compact range $[0, U_A+U_B]$ of $U_M(\cdot)$ by Lemma~\ref{lem:classC}(c), with no additional hypothesis; and $\psi$ is Lipschitz by assumption and bounded by $1$, while $f$ is bounded by $f(U_A+U_B)$ on the relevant range. The product and composition $s = f(U_M(\cdot))\,\psi(L_M(\cdot),S_M(\cdot),G_M(\cdot)) - \mu_A - \mu_B$ of bounded Lipschitz functions is therefore Lipschitz on $(0,+\infty)$; let $\Lambda$ be its constant.

\emph{Step 2 (right half-width).} Note first that $\varepsilon_+ = r_* s_0/(\mu_B + r_*\Lambda) < s_0/\Lambda$ and $\varepsilon_- < r_* s_0/(r_*\Lambda) = s_0/\Lambda$, so throughout Steps 2--3 we have $\Lambda\varepsilon < s_0$, hence $s(r) \geq s_0 - \Lambda\varepsilon > 0$ and $\mu_A + s(r) > \mu_A > 0$: the band form \eqref{eq:wealthband} is applicable at every rate considered. Now let $\varepsilon \in (0, \varepsilon_+]$ and $r = r_* + \varepsilon$. The upper inequality in \eqref{eq:wealthband} at $r$ is implied by
\[
r_* + \varepsilon \;\leq\; r_* + \frac{r_*(s_0 - \Lambda\varepsilon)}{\mu_B}
\quad\Longleftrightarrow\quad
\varepsilon\,(\mu_B + r_*\Lambda) \;\leq\; r_* s_0,
\]
which holds by the definition of $\varepsilon_+$. The lower inequality in \eqref{eq:wealthband} holds at $r > r_*$ a fortiori: since $s(r) > 0$, its left-hand side is $r_*\mu_A/(\mu_A + s(r)) < r_* < r$.

\emph{Step 3 (left half-width).} First, $\varepsilon_- < r_*$ (as $s_0 < \mu_A + s_0 + r_*\Lambda$), so $r = r_* - \varepsilon > 0$ for $\varepsilon \in (0,\varepsilon_-]$, and the upper inequality in \eqref{eq:wealthband} holds a fortiori at $r < r_*$: since $s(r) > 0$ (Step 2), its right-hand side exceeds $r_* > r$. For the lower inequality, using $s(r) \geq s_0 - \Lambda\varepsilon$ and the fact that $t \mapsto \mu_A/(\mu_A + t)$ is decreasing, it suffices
\[
(r_* - \varepsilon)\big(\mu_A + s_0 - \Lambda\varepsilon\big) \;\geq\; r_*\mu_A.
\]
Expanding, the left side equals $r_*\mu_A + r_*(s_0 - \Lambda\varepsilon) - \varepsilon(\mu_A + s_0 - \Lambda\varepsilon) \geq r_*\mu_A + r_* s_0 - \varepsilon\,(\mu_A + s_0 + r_*\Lambda)$, so the display holds whenever $\varepsilon(\mu_A + s_0 + r_*\Lambda) \leq r_* s_0$, i.e.\ for all $\varepsilon \leq \varepsilon_-$.

(iii) Each of the four defining functions for adoption, security, governance and liquidity preservation is continuous at $r_*$ ($\tau_A,\tau_B$ by Assumption~\ref{ass:migration}; 
$S,G,L$ by Proposition~\ref{prop:intervals}; $s$ by Proposition~\ref{prop:sbounded}) and satisfies its defining inequality \emph{strictly} at $r_*$ under strict admissibility and 
$s_0>0$. For each condition separately, continuity therefore provides $\varepsilon_j>0$ such that the inequality persists on $[r_*-\varepsilon_j, r_*+\varepsilon_j]$.  For the 
wealth condition an explicit admissible $\varepsilon_j$ is computed in (ii), so that taking $\varepsilon := \min_j \varepsilon_j > 0$ over the five conditions gives 
$[r_*-\varepsilon,r_*+\varepsilon] \subseteq \mathcal{R}$.
\end{proofof}

\end{document}